\documentclass{article}

\usepackage{arxiv}
\usepackage[utf8]{inputenc} 
\usepackage[T1]{fontenc}    
\usepackage{hyperref}       
\usepackage{url}            
\usepackage{booktabs}       
\usepackage{amsfonts}       
\usepackage{nicefrac}       
\usepackage{microtype}      
\usepackage{graphicx}
\graphicspath{{./images/}}  
\usepackage{siunitx}
\usepackage{array}
\usepackage{stfloats}
\newcolumntype{P}[1]{>{\centering\arraybackslash}m{#1}}
\newcolumntype{M}[1]{>{\raggedright\arraybackslash}m{#1}}
\usepackage{tabularray}

\title{Demonstrating a family of X-ray dark-field retrieval approaches on a common set of samples}

\author{
 Samantha J. Alloo \\
  School of Physics and Astronomy, Monash University, Australia \\
  \texttt{samantha.alloo@monash.edu} \\
  \And
 Ying Ying How \\
  School of Physics and Astronomy, Monash University, Australia \\
  Munich Institute of Biomedical Engineering, Technical University of Munich (TUM), Germany \\
  Research Group Biomedical Imaging Physics, Department of Physics, TUM School of Natural Sciences, TUM, Germany \\
  Chair of Biomedical Physics, Department of Physics, TUM School of Natural Sciences, TUM, Germany \\
  \And
 Jannis N. Ahlers \\
  School of Physics and Astronomy, Monash University, Australia \\
  \And
 David M. Paganin \\
  School of Physics and Astronomy, Monash University, Australia \\
  \And
 Michelle K. Croughan \\
  School of Physics and Astronomy, Monash University, Australia \\
  \And
 Kaye S. Morgan \\
  School of Physics and Astronomy, Monash University, Australia \\
}

\begin{document}
\maketitle


\begin{abstract}
Sensitive to scattering from unresolved sample structures, the dark-field channel in full-field X-ray imaging provides complementary information to that offered by conventional attenuation and phase-contrast methods. A range of experimental dark-field techniques and retrieval algorithms have been recently developed to extract this signal by directly resolving dark-field-associated local image blurring with a high-resolution camera. While the underlying physical mechanism that generates dark-field contrast is generally defined similarly across the methods, no comparison of these dark-field techniques using identical samples has been conducted. In this paper, dark-field imaging data from two samples were acquired using three emerging dark-field setups at a synchrotron: propagation-based, single-grid, and speckle-based X-ray imaging. Dark-field images were then reconstructed using a variety of retrieval algorithms—some requiring only a single sample exposure, others multiple; some performing local, pixel-wise analysis, and others operating globally on entire images. We find that the dominant contribution to dark-field contrast---arising from diffuse scattering from unresolved microstructures or multiple refractions through larger, potentially resolved structures---is consistently recovered across all approaches, demonstrating mutual agreement. However, some differences emerge for structures that are spatially varying, such as sharp edges. We attribute these differences to the idea that each technique has a different \textit{internal ruler}, a sensitivity scale for dark-field retrieval influenced by both the experimental design and algorithmic assumptions of the technique. We discuss the origin of these internal rulers in the context of the approaches examined. This study is intended to guide dark-field imaging users in selecting the most appropriate technique for their imaging goals and to motivate future research into dark-field sensitivity and sources of dark-field contrast across different methods.
\end{abstract}

\section{Introduction}
\begin{figure}
    \centering
    \includegraphics[width=0.65\textwidth]{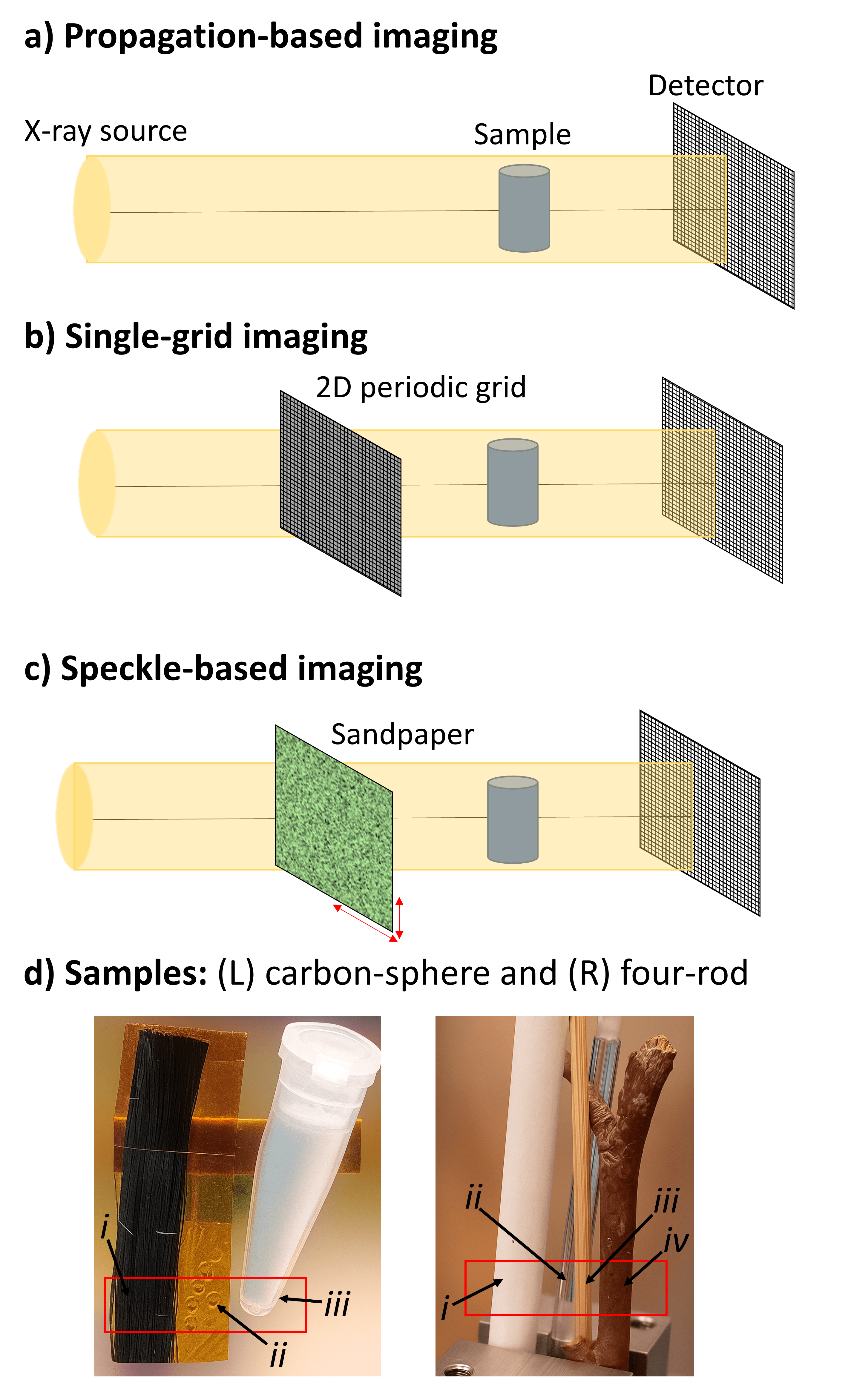}
    \caption{Experimental setups for the dark-field (DF) imaging techniques explored in this study: a) Propagation-based imaging, b) single-grid imaging, which employs one two-dimensional periodic grid, and c) speckle-based imaging, which uses a sandpaper or random membrane. d) The two samples imaged using the setups in a)--c), with the red boxes indicating the regions imaged. The carbon-sphere sample was comprised of \textit{i)} carbon fibers, \textit{ii)} $\qty{1.5}{\milli\meter}$ diameter polymethyl methacrylate (PMMA) spheres, and  \textit{iii)} $\qty{6}{\micro\meter}$ diameter PMMA microspheres in a plastic tube. The four-rod sample consisted of \textit{i)} a reed diffuser stick, \textit{ii)} a solid PMMA rod,  \textit{iii)} a toothpick, and \textit{iv)} a tree twig.
    } 
    \label{fig:setup} 
\end{figure}

\begin{figure}
    \centering
    \includegraphics[width=\textwidth]{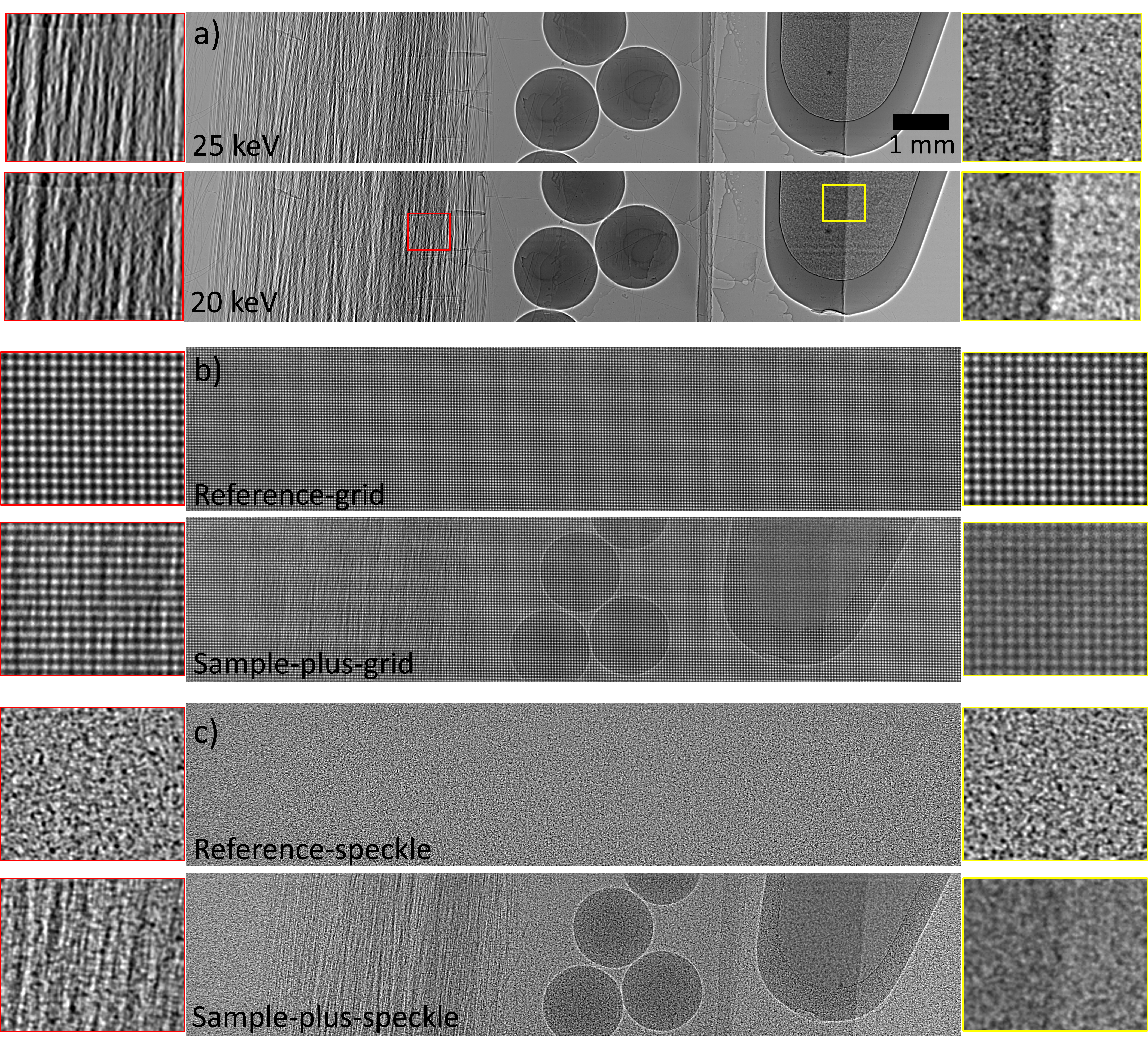}
    \caption{Images of the carbon-sphere sample, collected for the various DF approaches compared in this study. a)~Dual-energy propagation-based approach; DF contrast is generated by the difference in local blurring between two images collected at two sufficiently different X-ray energies \cite{ahlers2024x}. Note that the horizontal features that appear in the tube of microspheres at \qty{20}{\keV} originate from the flat-field correction. The illumination was not uniform, and the horizontal structures in that illumination were diffused by the DF effects generated by the microspheres, preventing the variations in the illumination from being divided out within the flat-field correction. b)~Sample-induced DF effect in the single-grid approach blurs the 2D-periodic reference-grid pattern \cite{how2022quantifying}. c)~In speckle-based imaging, DF contrast can be retrieved by tracking the speckle-blur between the reference-speckle and sample-plus-speckle images \cite{alloo2023m,beltran2023fast}. Note that in the case of the multi-speckle approach, images are collected for multiple positions of the speckle generator, but these are not shown here. } 
    \label{fig:data} 
\end{figure}

In the context of full-field X-ray imaging, scattering from unresolved, sub-resolution sample features gives rise to so-called dark-field (DF) effects \cite{Pagot2003,pfeiffer2008hard}. DF contrast is closely related to small-angle X-ray scattering (SAXS) and multiple refractions of the incident X-ray beam \cite{pfeiffer2008hard,yashiro2010origin,magnin2023dark}. However, sharp features in a sample, like edges, can also generate a DF signal \cite{yashiro2015effects, alloo2025separating}. DF-generating structures cannot typically be resolved using attenuation-based or phase-contrast imaging, but the scatter from these structures can be detected using specialist X-ray imaging setups, often the same setups as used for phase-contrast imaging. DF imaging involves capturing experimental X-ray intensity data on such setups and then applying an appropriate retrieval algorithm to extract the DF signal. The retrieval process typically not only yields a DF image, but also provides complementary attenuation and phase reconstructions--although this paper focuses solely on the DF retrieval aspect. When the retrieval algorithm incorporates DF effects, the retrieved attenuation and phase images are typically more accurate than when DF effects are omitted \cite{leatham2023x, ahlers2024x, alloo2024stabilizing}.
\\\\
There are several well-established X-ray DF imaging techniques, including grating interferometry \cite{pfeiffer2008hard}, analyzer crystal approaches \cite{Pagot2003,Wernick2003}, and edge-illumination \cite{Endrizzi2014}. Among the newer techniques are propagation-based \cite{Gureyev2020, leatham2023x, ahlers2024x}, single-grid \cite{Wen2010,morgan2013sensitive}, and speckle-based \cite{berujon2012x, zdora2018state} DF imaging, with experimental setups as summarised in Fig.~\ref{fig:setup}. These newer imaging setups all directly resolve DF effects by measuring local blurring in a raw experimental image, a blurring that is typically several microns in width, and hence are all well-suited to high-resolution imaging. With the aim of examining their similarities, differences, and individual strengths, this work investigates these emerging setups by capturing appropriate DF imaging data of two test samples and evaluating some of the relevant DF retrieval algorithms. The remainder of this introduction provides an overview of these DF imaging setups, along with the algorithms used for DF retrieval.
\\\\
Propagation-based imaging, shown in Fig.~\ref{fig:setup}a, requires no additional optics beyond the X-ray source, sample, and detector, although it necessitates a relatively high degree of spatial coherence at the sample \cite{Snigirev1995,Cloetens1996,Wilkins1996}. The distance between the sample and detector allows the coherent wavefield to self-interfere, resulting in bright and dark interference fringes that locally enhance edge contrast in the image of the sample. In this setup, DF effects from porous or granular regions in a sample result in locally reduced image contrast (or blur) compared to the image that would otherwise be seen. The resulting local image blur increases with decreasing X-ray energy \cite{ahlers2024x}, as lower-energy X-rays are scattered to larger angles, and with increasing propagation distance between the sample and detector \cite{leatham2023x}, as the scattered X-rays spread across more pixels. \\\\
Propagation-based imaging has been widely used for phase-contrast imaging \cite{gureyev2009refracting,topperwien2018three}, and has only recently been employed to capture DF information \cite{Gureyev2020, leatham2023x, ahlers2024x}. Recent methods separate DF from phase-contrast effects by acquiring two propagation-based images under conditions where the contributions from these effects differ significantly. For example, Leatham~\textit{et al.}~varied the propagation distance, while Ahlers~\textit{et al.}~used two different X-ray energies. In this work, we retrieve the DF signal using the approach in Ahlers~\textit{et al.} from two propagation-based X-ray imaging data sets acquired at two different monochromatic energies. Suitable data for this approach are shown in Fig.~\ref{fig:data}a, where propagation-based images of one of our test samples have been captured at energies of \qty{20}{\keV} and \qty{25}{\keV}. DF effects are stronger in the lower-energy image, as indicated by the increased image blurring behind strongly scattering regions, which are highlighted in the red and yellow magnified regions within this figure. The dual-energy propagation-based DF retrieval approach we use is derived from the Fokker--Planck equation for X-ray imaging, an equation that models coherent (phase shift) and diffusive (DF effects) flows of intensity in an X-ray imaging context \cite{paganin2019x, morgan2019applying}. We employ the single-material local DF retrieval approach described in Ahlers~\textit{et al.}, where the first step is to use the X-ray Fokker--Planck equation to recover the sample thickness using both experimental images. The simultaneous consideration of phase and DF effects in this step means that the quality of the retrieved thickness is better than what it would be if the blurring DF effects were ignored. The retrieved sample thickness is then used to simulate a so-called DF-free image, and the DF signal is then recovered by locally comparing the image visibility in the DF-free image to that in the original lower-energy experimental image. The Fokker--Planck DF coefficient is then calculated from the visibility using the propagation distance and period of local intensity variations. This DF imaging approach is referred to as `self-referencing' because it requires no external beam modulation. However, it does rely on the sample containing sufficient texture that the resulting image contrast can be visibly blurred at the lower imaging energy. 
If the sample lacks sufficient texture within an analysis window, the measured visibility is dominated by noise, which is then amplified by the ratio of visibilities. To address this, a small, manually optimized regularization parameter can be introduced, which effectively suppresses this noise (see Table~1 in the supplementary materials for details).
\\\\
Single-grid imaging inserts a modulator into the setup for propagation-based imaging placed immediately before or after the sample (see Fig. \ref{fig:setup}b 
for the setup in which the modulator is upstream of the sample). In this technique, the modulator is used to generate a 2D-periodic grid pattern in the X-ray illumination. The modulation grid either attenuates \cite{Wen2010, morgan2011, how2022quantifying} or phase-shifts \cite{morgan2013sensitive,rizzi2013} the incident illumination to generate the periodic reference pattern, with the latter minimally attenuating the X-rays, thereby improving imaging speed.  The modulation in the imaging illumination enables the detection of sample-induced gradients in the X-ray phase, increasing sensitivity to subtle variations in the sample thickness or density relative to propagation-based imaging. While local shifts in the grid pattern reveal the phase of the X-ray wavefield, local blurring of the pattern reveals a DF effect. The well-defined periodic illumination in single-grid imaging enables controlled tuning of the system's sensitivity to DF effects by adjusting the modulator-to-detector distance and/or the modulator, and hence the period of the generated reference pattern. This tunability allows quantitative analysis of the recovered DF signal, as the measured DF can be correlated with the number of structures generating that signal \cite{lim2022quantification,how2022quantifying,how2023}. Periodic grids provide retrieved DF images with uniform spatial resolution and sensitivity across the entire field of view. This means that the imaging system is most sensitive to a single scattering signal, as determined by the grid period, while still having certain sensitivity towards scattering signals of other length scales comparable to the grid period. Single-grid imaging is however vulnerable to Moir\'e artefacts if the sample contains periodic features similar in period to the grid pattern. Artefacts resembling the grid pattern may also arise in reconstructed images as a result of the varying sensitivity and the sampling process used for DF retrieval. In such cases, applying a post-processing median filter with a kernel size equal to the grid period, or an appropriate Fourier filter, can help suppress these artefacts.\\\\
To retrieve a DF image using a single-grid setup, a reference-grid image and a sample-plus-grid image must be captured. The reference-grid image is captured with just the grid in the beam, and the sample-plus-grid image is captured when the sample is inserted into the grid-modulated beam. Figure~\ref{fig:data}b shows a pair of these images for one of the samples investigated in this work. In regions with dense unresolved microstructure, the grid pattern will be locally blurred or reduced in visibility, for example, behind the tube of microspheres shown in the yellow box in Fig.~\ref{fig:data}b. A sample with unresolved fibers or orientated structures will predominantly blur the grid pattern in the direction perpendicular to those fibers, as seen in the red box in Fig.~\ref{fig:data}b, revealing a directional DF signal \cite{croughan2023directional}. A sample's DF image can be recovered by mathematically tracking this sample-induced blurring using an appropriate retrieval algorithm. In this study, we apply How and Morgan's single-exposure DF approach, which uses local pixel-wise cross-correlation analysis to quantify the difference in visibility between the reference and sample-plus-grid images \cite{how2022quantifying}. Specifically, the periodic grid intensity patterns are modeled as a sinusoid, and the cross-correlation is performed within local analysis windows (typically sized to match the grid period) in both the reference-grid and sample-plus-grid images to measure the differences in this model with and without the sample present. The explicit modeling and tracking of intensity changes imposed by the sample means that phase and DF effects are simultaneously considered and thus decoupled, within limitations imposed by the choice of cross-correlation parameters. Readers are referred to How and Morgan's paper for specific details regarding the derivation and implementation of the algorithm \cite{how2022quantifying}, and Croughan \textit{et al.}'s paper for a directional DF generalization of the algorithm \cite{croughan2023directional}. 
\\\\
Speckle-based X-ray imaging, shown in Fig.~\ref{fig:setup}c, replaces the periodic modulator in single-grid imaging with a spatially random one. This inherently alleviates the need for high precision in the mask fabrication, to the extent that a simple sheet of sandpaper \cite{morgan2012} or a biological membrane \cite{berujon2012x} can be used as a modulator. A reference-speckle and sample-plus-speckle image pair must be recorded to capture and then retrieve the DF signal---an example data pair is shown in Fig.~\ref{fig:data}c. Similarly to the case of single-grid imaging, DF effects manifest as speckle blurring or a reduction in visibility in the sample-plus-speckle image compared to the reference-speckle image. Some DF retrieval approaches for the speckle-based imaging require a single pair of data, while others require multiple pairs acquired by moving the modulator to different transverse positions---either equidistant or random, depending on the specific retrieval algorithm---and collecting a reference-speckle and a sample-plus-speckle image pair at each step \cite{zdora2018state}. 
The use of multiple pairs can improve the spatial resolution of the retrieved images. Due to the random nature of the reference-speckle pattern, the sensitivity across the system's field of view can vary. For example, speckle patterns with regions containing large speckles exhibit reduced sensitivity. Large speckles are more difficult to shift (phase effects) or blur (DF effects) than smaller ones, so changes induced by the sample are subtle and can result in the image retrieval algorithm experiencing difficulties in these regions. This issue can be somewhat alleviated in multiple-pair speckle-based imaging, since a smaller speckle is likely to pass over a given area in a subsequent image pair that will then contribute to the final reconstruction. The variation in speckle size in speckle-based imaging can be advantageous for samples with hierarchical structures, enabling sensitivity to a broad range of DF scattering signals corresponding to the range of characteristic length scales in the speckle pattern \cite{meyer2021quantitative}. \\\\
In this work, we explore two DF algorithms for speckle-based imaging: one that requires a single image pair \cite{beltran2023fast} (referred to as single-speckle imaging hereafter) and one that requires multiple pairs \cite{alloo2023m} (referred to as multi-speckle imaging hereafter). Like the dual-energy propagation-based DF retrieval method described earlier, the speckle-based DF retrieval algorithms used in this work are derived from the X-ray Fokker--Planck equation. However, speckle-based approaches are global methods, as the DF is retrieved by evaluating an analytically derived equation based on the relevant Fokker--Planck formalism using entire experimental images. Another difference between the dual-energy propagation- and speckle-based DF retrieval algorithms is that the former strictly assumes a single material, requiring \textit{a priori} information of both components $\delta$ and $\beta$ of a sample's complex refractive index decrement, where $n = 1 - \delta + i\beta$. The single- and multi-speckle-based approaches, on the other hand, assume the sample is homogeneous, meaning that $\gamma = \delta/\beta$ is constant, requiring \textit{a priori} information of this $\gamma$ value for successful DF retrieval. This assumption enables the speckle-based approaches to remain valid for multi-material samples, where constituent materials have different refractive indices, provided that the $\delta$/$\beta$ ratio value is approximately the same across all materials. In contrast, the single-material assumption in Ahlers~\textit{et al.}'s dual-energy propagation-based approach may be less reliable when the sample contains materials with significantly different attenuation and refraction characteristics, as it assumes $\delta$ and $\beta$ are individually constant throughout the sample. Besides the homogeneous sample assumption, other assumptions made by the two speckle-based DF retrieval methods investigated here are fundamentally different, aspects of which we briefly discuss.
\\\\
\noindent In Beltran~\textit{et al.}'s single-speckle DF imaging method, the DF is retrieved by first obtaining an approximation of the phase. This phase retrieval step assumes that the DF effects are weak compared to the attenuation and propagation-based phase effects, allowing a sample-only image to be estimated by dividing the sample-plus-speckle image by the reference-speckle image. This is a reasonable assumption if the reference speckles are not significantly blurred or shifted by the sample, and allows a conventional transport-of-intensity-based phase retrieval to be applied \cite{paganin2002}. The transport-of-intensity-equation-retrieved phase is then substituted into a form of the Fokker--Planck equation for speckle-based imaging, which is solved to obtain the DF image. The two-step nature of this method, in which DF effects are neglected in the first step, means that phase and DF effects are not considered simultaneously. However, as Beltran~\textit{et al.} suggested, the initial phase approximation can be updated to account for DF effects using an iterative extension of the method, which we do not implement in this work. \\\\
If several imaging acquisitions of the sample are possible, pairs of reference-speckle and sample-plus-speckle images can be captured, and a multi-exposure DF retrieval algorithm can improve the quality of the retrieved images. Here, we explore the multi-speckle DF approach in Alloo~\textit{et al.} \cite{alloo2023m}, the most generalized variant of the so-called Multimodal Intrinsic Speckle-Tracking (MIST) algorithm \cite{pavlov2020x}. By using multiple data pairs for image retrieval, the approach simultaneously retrieves phase and DF images, leveraging the additional data to disentangle DF and phase-contrast effects. Specifically, it requires a minimum of four speckle imaging pairs, each of which establishes a linearized form of the Fokker--Planck equation for speckle imaging containing four unknown variables corresponding to the desired reconstructed images. Typical methods for solving a linear system can be applied; Alloo~\textit{et al.} suggest using Tikhonov-regularized QR decomposition. The phase and DF images are then computed using Fourier-space filtering to suitably combine the solutions from the Tikhonov-regularized QR decomposition. Additional data above the minimum of four speckle-imaging pairs can further improve image quality, as the linear system to be solved becomes overdetermined, thereby increasing redundancy in the data.
\\\\
In this paper, we aim to explore similarities and differences between propagation-based, single-grid, and speckle-based setups for DF imaging, as well as the associated DF retrieval algorithms described earlier in this introduction. Note that the data sets for the single-grid and single-speckle retrieval algorithms can be exchanged, but we focus on one retrieval approach for each setup here. In the research literature, there have been review studies that compare different experimental phase-contrast techniques \cite{zhou2008development,wilkins2014evolution,endrizzi2018x, ruiz2016}, as well as studies that compare retrieval algorithms for one specific experimental technique, for example, in speckle-based imaging \cite{rouge2021comparison,zandarco2024speckle,celestre2024review}. As far as we are aware, this is the first study to compare different DF imaging techniques on the same sample. We demonstrate mutual agreement between the retrieved DF images of two samples across the four methods (dual-energy propagation-based, single-grid, single-speckle, and multi-speckle imaging), as well as the complementarity of the different techniques and algorithms. The remainder of the paper is organized as follows. Section~\ref{experimental} describes the experimental data collection procedures for the various X-ray DF techniques examined in this work. Section~\ref{results} presents the retrieved DF images obtained using different retrieval algorithms, highlighting shared features and distinctive characteristics. In Sec.~\ref{summary}, we summarize our findings and provide a user-friendly table comparing the DF imaging setups and retrieval algorithms explored.\\\\
\section{Experimental methods}\label{experimental}
\noindent We imaged two samples using all four DF methods; the samples, which we will refer to as the `carbon-sphere' sample and the `four-rod' sample, are shown in Fig.~\ref{fig:setup}d. Referencing the labels in Fig.~\ref{fig:setup}d, the carbon-sphere sample was comprised of \textit{i)} carbon fibers, \textit{ii)} $\qty{1.5}{\milli\meter}$ diameter polymethyl methacrylate (PMMA) spheres, and  \textit{iii)} $\qty{6}{\micro\meter}$ diameter PMMA microspheres in a plastic tube. The four-rod sample consisted of \textit{i)} a reed diffuser stick, \textit{ii)} a solid PMMA rod,  \textit{iii)} a toothpick, and \textit{iv)} a tree twig. Imaging was performed in the first experimental hutch of the MicroCT beamline at the Australian Synchrotron, which is located \qty{24}{\meter} from the bending-magnet source point. The detector system for all techniques was a pco.edge 5.5 complementary metal-oxide-semiconductor (CMOS) camera, which had $2560 \times 2160$ pixels with a \qty{6.5}{\micro\meter} $\times$ \qty{6.5}{\micro\meter} pixel size, coupled to a GGG:Eu/Tb scintillator with a 1$\times$ optical lens placed in between, and 0.04 s exposure times were used throughout. For the dual-energy propagation-based approach, images at monochromatic X-ray energies of \qty{20}{\keV} and \qty{25}{\keV} were acquired with a fixed propagation distance of \qty{0.7}{\meter} between the sample and the detector. For both wavefront-modulated techniques, single-grid and speckle-based imaging, the wavefront modulator was positioned \qty{0.3}{\meter} in front of the sample, the sample-to-detector propagation distance was \qty{0.7}{\meter}, and a \qty{25}{\keV} monochromatic X-ray beam was used. A geological stainless steel sieve with $\qty{25}{\micro\meter}$ square apertures and $\qty{27}{\micro\meter}$ thick wires, producing a $\qty{52}{\micro\meter}$ period grid pattern at the detector, was used for single-grid imaging. For speckle-based imaging, two layers of grit P800 sandpaper were inserted in place of the grid. The generated speckle pattern at the detector had an average speckle size of $\qty{18.3}{\micro\meter}$, measured by taking the full-width at half maximum of the reference-speckle pattern's autocorrelation function \cite{goodman2020speckle}. To capture multi-speckle imaging data of both samples, the speckle mask was moved to 13 random transverse positions, collecting 13 pairs of reference-speckle and sample-plus-speckle images. For all imaging methods, dark-current and flat-field images were acquired to correct for detector electronic noise and X-ray beam inhomogeneities.

\section{Results and discussion} \label{results}
DF images of the carbon-sphere and four-rod samples were retrieved using the approaches described in the introductory text of this paper: dual-energy propagation-based imaging, single-grid imaging, single-speckle imaging, and multi-speckle imaging. The input parameters used for DF retrieval within each approach are listed in the supplementary material.

\subsection{Carbon-sphere sample}
\begin{figure}
    \centering
    \includegraphics[width=0.7\textwidth]{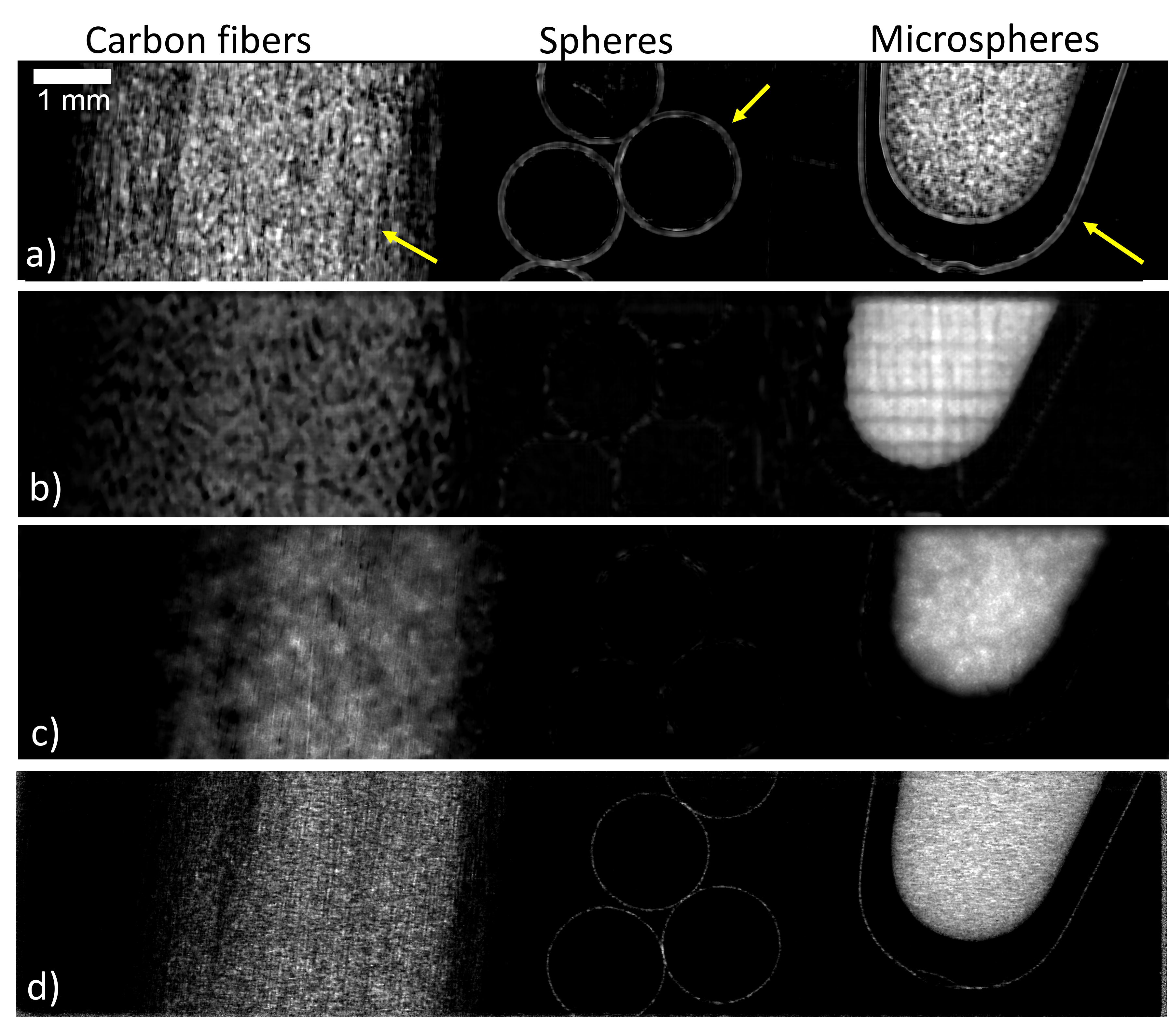} 
    \caption{Retrieved DF images of the carbon-sphere sample using the different retrieval approaches: a) dual-energy propagation-based, b) single-grid, c) single-speckle, and d) multi-speckle. The grayscale ranges for the images are [min (black), max (white)]: a)~=~[0.0, 1.3]$\times10^{-10}$, b)~=~[0.0, 0.5], c)~=~[0.0, 4.7]$\times10^{-11}$, and d)~=~[0.0, 5.8]$\times10^{-11}$.}
    \label{fig:CarSpher}
\end{figure}

\noindent Figure~\ref{fig:CarSpher} shows the retrieved DF images of the carbon-sphere sample using the four different methods. At first glance, the DF image in Fig.~\ref{fig:CarSpher}d obtained using the multi-speckle approach looks the best, exhibiting characteristics typically associated with a high-quality image: low background noise, sharp features, and minimal artefacts. This is expected as this method incorporates the largest data set: 13 pairs of speckle-based images, meaning that more information about the sample is encoded in the data, supporting high-quality DF image retrieval. The DF images in Figs.~\ref{fig:CarSpher}b and \ref{fig:CarSpher}c rely on a single sample image, while the image in Fig.~\ref{fig:CarSpher}a uses two. In this context, it is noteworthy that these other, much lower-exposure methods---dual-energy propagation-based, single-grid, and single-speckle---retrieve DF images with characteristics broadly comparable to those obtained using the multi-speckle approach, which relies on significantly more data. In the multi-speckle approach, acquiring pairs of speckle-based imaging data at different mask positions probes the sample in slightly different ways, with each data set encoding additional information. It follows that a reconstructed signal that incorporates more data sets would yield higher image quality than one retrieved from a single data set. This improvement could be expected in two image quality metrics -- 1) higher spatial resolution, since the single-exposure methods extract measurements from a neighborhood of pixels, not a single pixel, and 2) higher DF sensitivity because multiple measurements typically increase robustness to noise. While single-exposure DF retrieval approaches generally exhibit lower image quality than approaches using multiple data sets, they offer the important advantage of enabling dynamic imaging \cite{croughan2023directional,how2024vivo, smith2024}, which is not currently feasible with multi-exposure techniques (except in the case of repeated motion \cite{gradl2018}). On this note, the dual-energy propagation-based approach is also capable of dynamic imaging if a dual-channel photon-counting detector is used, as the two required data sets can then be captured simultaneously \cite{ahlers2025single}.
\\\\
The influence of both the reference pattern's sensitivity to sample features and the number of data sets used for retrieval is realized by comparing the PMMA microsphere sample on the right of the DF images in Fig.~\ref{fig:CarSpher}. The $\qty{6}{\micro\meter}$ diameter PMMA microspheres are smaller than the detector's $\qty{6.5}{\micro\meter}$ pixel size, and are therefore expected to contribute to the DF signal arising from diffuse X-ray scattering on unresolved microstructure. The PMMA microspheres are predicted to generate a relatively slowly-varying DF signal due to their sub-pixel size and the slowly-varying sample thickness in this region. However, the DF retrieved using the single-speckle approach in Fig.~\ref{fig:CarSpher}c appears somewhat inhomogeneous. This discrepancy may be attributed to the single measurement of the carbon-sphere sample in this approach, where the sampling efficacy is strongly influenced by speckle pattern characteristics such as size and sparsity. In contrast, when a periodic reference pattern is used, DF retrieval tends to be more stable, as the sample is probed consistently across the entire image, as seen in Fig.~\ref{fig:CarSpher}b, although some artefacts from the reference pattern and/or horizontal structures in the illumination remain. Finally, although the dual-energy propagation-based approach relies solely on energy-associated differences in DF blur of the sample image itself and does not use an introduced reference pattern, it nevertheless shows remarkable agreement with the alternative approaches.
\begin{figure}
    \centering
    \includegraphics[width=0.7\textwidth]{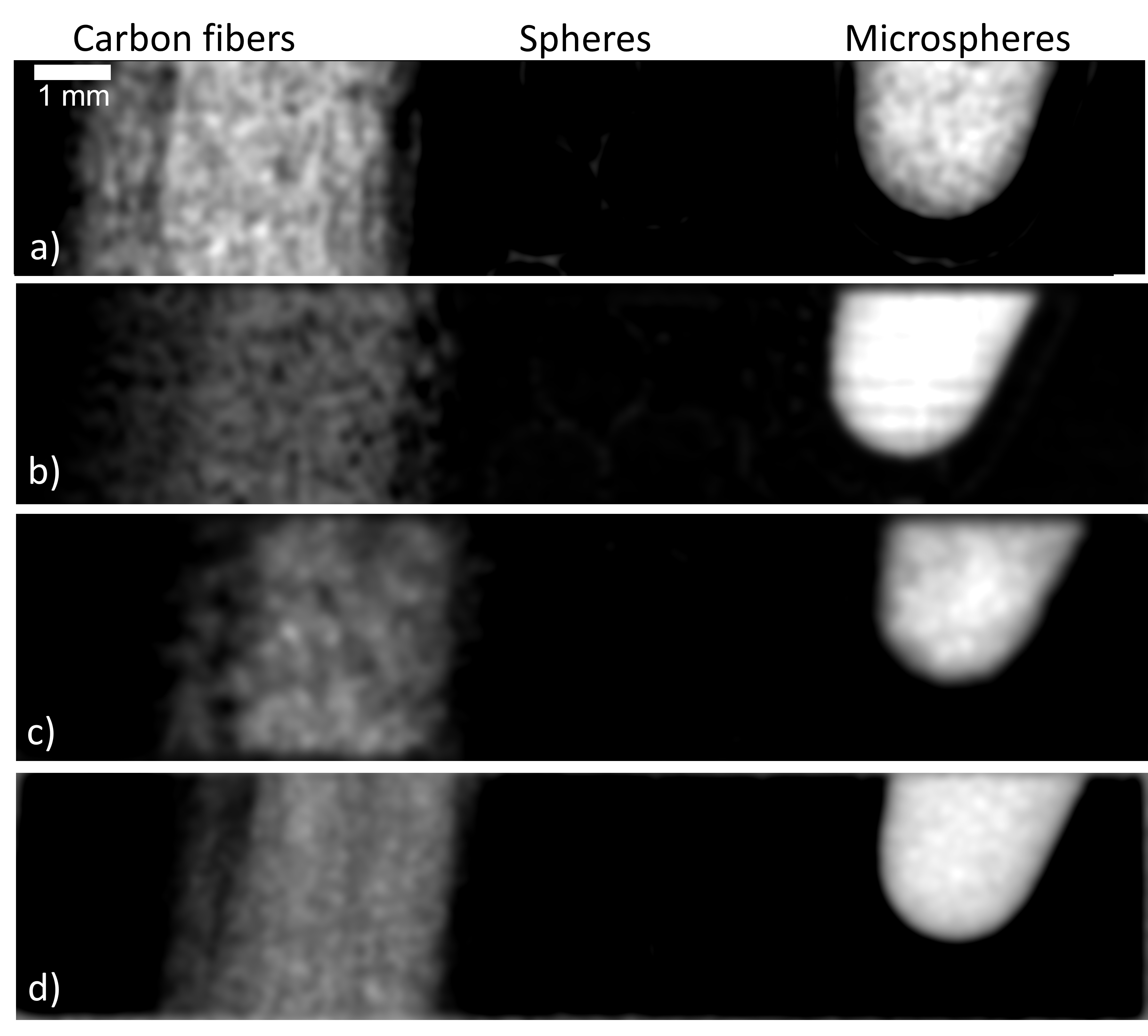} 
    \caption{The retrieved DF images shown in Fig.~\ref{fig:CarSpher} (with the same labeling of sub-figures) with a 10-pixel standard deviation Gaussian filter applied. The approaches used to retrieve each of the images were: a) dual-energy propagation-based, b) single-grid, c) single-speckle, and d) multi-speckle. The grayscale ranges for the images are [min (black), max (white)]: a)~=~[0.0, 9.7]$\times10^{-11}$, b)~=~[0.0, 0.4], c)~=~[0.0, 4.0]$\times10^{-11}$, and d)~=~[0.0, 4.7]$\times10^{-11}$.}
    \label{fig:CarSpher_blur}
\end{figure}
\\\\
\noindent To compare the low-spatial-frequency similarities in the recovered DF images of the carbon-sphere sample, a 10-pixel standard deviation Gaussian filter was applied to all DF images in Fig.~\ref{fig:CarSpher}, yielding those shown in Fig.~\ref{fig:CarSpher_blur}. Notably, the Gaussian-filtered variants appear particularly similar across the different DF retrieval approaches: both the bundle of carbon fibers and tube-contained PMMA microspheres produce a clearly retrievable DF signal, which are on the left and right of the sample, respectively, while the larger, solid spheres in the middle generate no DF signal. This indicates that the DF images share similar global characteristics and are broadly consistent in their low spatial frequencies.\\\\
The relative strength of the DF signals within each image is also of interest. In the dual-energy propagation-based DF image, the retrieved DF signals from the carbon fiber and the microsphere samples are similar in strength, as seen most clearly in the blurred variant of the retrieved DF in Fig.~\ref{fig:CarSpher_blur}a. In contrast, the other three reconstructions show a noticeably stronger DF signal from the PMMA microspheres than from the carbon fiber bundle. This disparity may be due to the different X-ray energies used for the imaging measurements -- the dual-energy approach primarily extracts local blurring within the \qty{20}{\keV} image (using the \qty{25}{\keV} in establishing a reference), while the other three approaches measure at \qty{25}{\keV}. The DF signal measured from the microspheres may be approaching saturation in the experimental images at the lower energy; hence, the difference between the carbon fibers and the microspheres is not as clear. Such DF sensitivity in the dual-energy propagation approach can be adjusted by changing the energy, or when a modulator is used, the period of the reference pattern can be chosen to adjust the sensitivity and can be specified independently of the sample. Figures~\ref{fig:CarSpher}b--d suggest that the single-grid and speckle-based systems' configurations (modulation size, propagation distance, and X-ray energy) allowed the resulting DF effects from the microspheres and carbon fibers to differ sufficiently, allowing clear differentiation between these two materials. \\\\
The methods explored produce markedly different DF reconstructions at sharp sample features, such as edges. These features are comparable in size to the system's spatial resolution and produce strong phase-Laplacian effects. Examples of these structures are indicated by yellow arrows in Fig.~\ref{fig:CarSpher}a and include the edge of the plastic tube containing the PMMA microspheres and the edges of the larger spheres, visible in Figs.~\ref{fig:CarSpher}a, b, and d, as well as the individual carbon fibers within the bundle, visible in all except the single-grid-retrieved DF image (Fig.~\ref{fig:CarSpher}b). Edges generate DF contrast through a mechanism distinctly different from the diffuse scattering/SAXS and multiple refraction contributions that are typically associated with X-ray DF contrast. How this edge-associated DF manifests in imaging data acquired using different experimental techniques and how it is treated within various DF retrieval approaches remains a relatively open question in the current research literature \cite{alloo2025separating}. 
\subsection{Four-rod sample}
\begin{figure}
    \centering
    \includegraphics[width=0.7\textwidth]{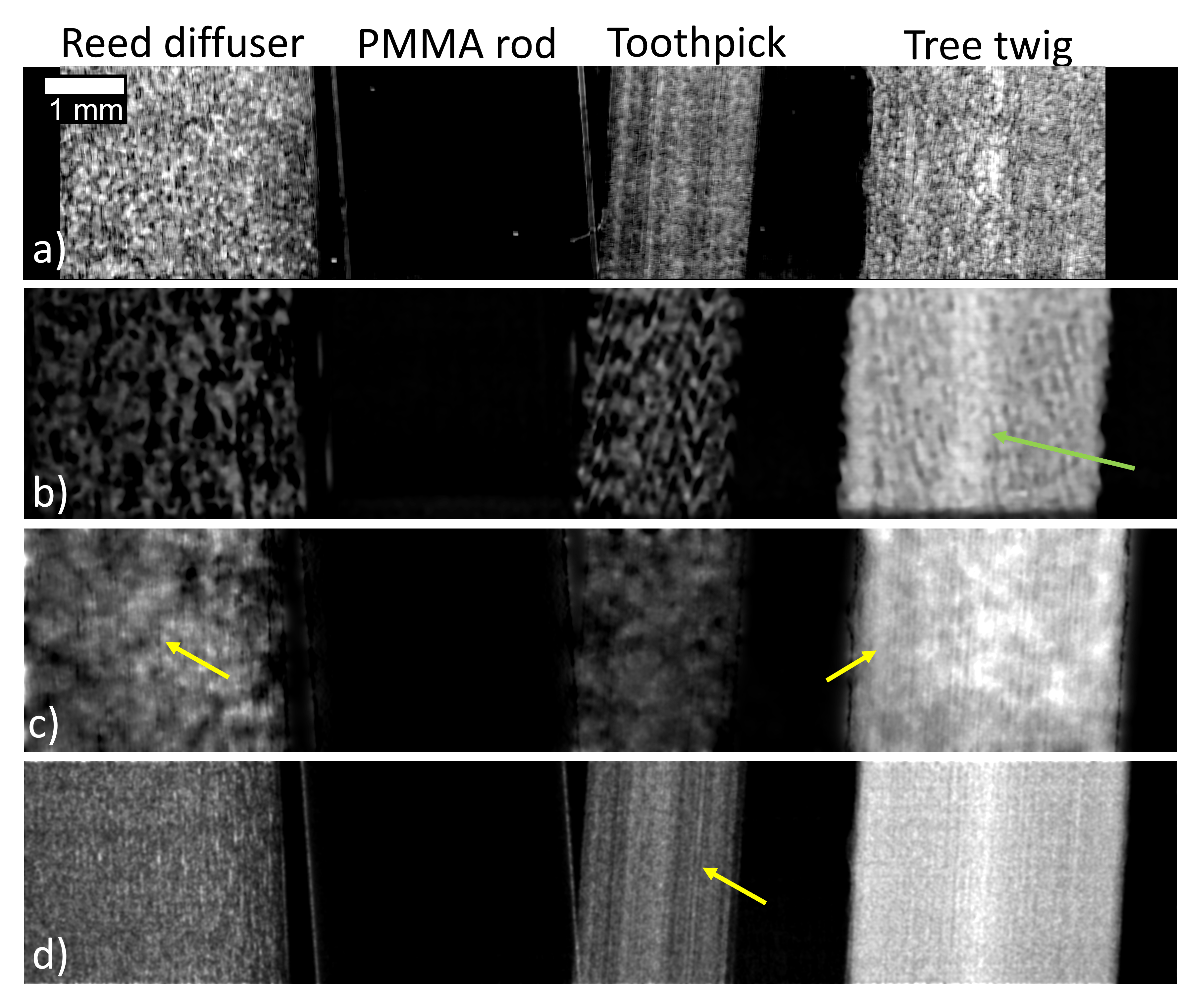} 
    \caption{Retrieved DF images of the four-rod sample using the different retrieval approaches, with a 2-pixel standard deviation Gaussian filter applied post-retrieval: a) dual-energy propagation-based, b) single-grid, c) single-speckle, and d) multi-speckle. The grayscale ranges for the images are [min (black), max (white)]: a)~=~[0.0, 1.4]$\times10^{-10}$, b)~=~[0.0, 0.6], c)~=~[0.0, 4.1]$\times10^{-11}$, and d)~=~[0.0, 4.1]$\times10^{-11}$.}
    \label{fig:FourRod}
\end{figure}

\noindent The DF images of the four-rod sample using the various retrieval methods are shown in Fig.~\ref{fig:FourRod}. Similar trends to those observed for the carbon-sphere sample are evident here: there is mutual agreement in the retrieved DF in microstructure-dense regions that give rise to a low-spatial-frequency DF signal; for example, the cortex (outer region) of the tree twig. However, differences are evident in both (1) the relative DF contrast between materials and (2) the resolvability of fiber-like structures. \\\\
The tree twig on the right-hand side of the four-rod sample shows the strongest DF signal. Moreover, all retrieval methods agree on its relative contrast compared to the other rods. However, only the single-grid and multi-speckle approaches clearly distinguish between the twig's inner pith (indicated by the green arrow in Fig.~\ref{fig:FourRod}b) and cortex, with the former producing a stronger signal than the outer cortex region. This feature is faintly visible in the dual-energy-retrieved DF (Fig.~\ref{fig:FourRod}a) but is obscured by other textured structures within the twig’s DF signal obtained using this method. It is absent in the single-speckle result in Fig.~\ref{fig:FourRod}c, likely because this retrieval method is highly sensitive to the characteristics of the reference speckle pattern; in this case, bright speckles over the central region of the twig overcast and mask the feature.
\\\\
Clear differences emerge across retrieval methods when comparing the retrieved DF in the toothpick and reed diffuser using the different methods. The dual-energy propagation-based and single-speckle approaches agree that the reed diffuser generates a stronger DF signal than the toothpick. In contrast, the single-grid and multi-speckle approaches retrieve these signals as broadly comparable, with the latter revealing slightly more structural details as the former suffers from artefacts. 
\\\\
\noindent The observed differences in retrieved DF contrasts across the methods may result from a trade-off between a method’s sensitivity range and its precision in DF retrieval. Broadly speaking, \textit{range} refers to the breadth of a method's sensitivity, or the span of DF strengths it can detect. A method with a broad DF range can capture contributions from both weakly and strongly scattering regions of a sample in the same measurement, revealing differences in DF magnitude between them. In contrast, the \textit{precision} of a DF method refers to the fineness and repeatability of its measurements, namely its ability to accurately distinguish small differences in the DF signal. As an analogy, range is how long your ruler is, while precision is how fine the markings are. \\\\
Considering the dual-energy propagation-based approach, the DF sensitivity and precision are both dependent on the local contrast or texture the sample introduces to the experimental image and how significantly the DF effects differ between the two energies, that is, how much the local image blurring differs. If the difference is such that the texture becomes visibly blurred at the lower energy compared to the higher, yet still resembles the original structure, then the DF can typically be retrieved. However, if this difference is either too small or too large, the sensitivity of DF retrieval may be reduced or artefacts may appear. The single-grid method tends to offer the highest precision due to its clearly defined, highly visible periodic reference pattern, which is most easily achieved using an absorbing grid. \\\\
The high-visibility periodic illumination makes this method most sensitive to blur widths (i.e., DF signals or characteristic scattering angles) on the order of the grid period. A key observation from the single-grid DF reconstruction of this sample is the presence of Moir\'e-like artefacts in the toothpick's retrieved DF, where the propagation-based phase-contrast fringes generated by its fibers are similar in size and contrast to the grid pattern. In addition, the image texture from the reed diffuser is similar to the grid pattern, resulting in areas where the reconstruction struggles. In regimes where the grid pattern is better separated from the sample-induced contrast (e.g., larger samples), this is less likely to be an issue.\\\\
Speckle-based imaging can likely avoid Moir\'e artefacts and access a broader range of blur widths (i.e., DF signals) due to greater variability in speckle characteristics across the detector's field of view, for example, differences in speckle visibility and size. This broader DF range enables the detection of a wider variety of structural differences. This helps explain why the structures inside the reed diffuser and toothpick are retrieved successfully in the DF using the multi-speckle approach in Fig.~\ref{fig:FourRod}d. The single-speckle approach suffers from limited access to a range of DF signal strengths, as sensitivity varies strongly across the image depending on the local speckle characteristics. Regions with large, sparse speckles show reduced sensitivity---evident by the speckle artefacts within the reed diffuser---whereas small, dense speckles provide increased sensitivity. This spatial variability is effectively mitigated in the multi-speckle approach. By sampling the illumination at multiple length scales via image acquisition at different mask positions, this method provides both high precision and sensitivity to a broad range of DF signals at the cost of increased sample exposure. The single-speckle method also accesses a broad DF range due to the natural variation in speckle size across the field of view, however, this comes at the cost of spatial resolution, particularly in regions where the reference-speckle pattern is sub-optimal. \\\\
A second difference between the retrieved DF images of the four-rod sample is the resolvability of fiber-like structures, indicated by the yellow arrows in Fig.~\ref{fig:FourRod}. These structures lie on the cusp of being resolved in the sense that their characteristic length scales are approximately equal to, or just larger than, the spatial resolution of the imaging system. As a result, these structures contribute differently to the phase and DF signals across the various experimental techniques and image retrieval methods. The reed diffuser stick on the left of the four-rod sample demonstrates this effect. When captured in projection, the individual fibers of the diffuser stick span only a few pixels in width, placing them in the gray area between generating phase-contrast and/or DF-associated image blur \cite{pfeiffer2008hard,yashiro2010origin}. While this effect may be explained using the `range versus precision' argument given earlier in this manuscript, it can also be rationalized by considering the idea of \textit{internal rulers}. In effect, all of the trialed DF-retrieval approaches have different internal rulers, in the sense that each approach has a unique implicit sensitivity given by the length of said rulers. This is somewhat analogous to the correlation length used to quantify the sensitivity of X-ray grating interferometers \cite{lynch2011, strobl2014}. Here, this sensitivity is governed similarly by the wavelength, propagation distance, pixel size, as well as the visibility and period of the illumination from which blurring is measured. The length of these internal rulers determines which structures each contrast modality within a given technique is sensitive to. One can think: if an imaging system had a detector with infinitely small pixels, then all structures within a sample would be resolved, and no DF effects would appear in the acquired data. However, DF signal could still be retrieved due to limitations of the retrieval method -- highlighting the dependence of these internal rulers not only on the experimental setup but also on the theoretical underpinnings of the retrieval algorithm. On the other hand, if the pixels are sufficiently large, on the order of hundreds of microns, then many of the structures cannot be resolved and hence will contribute to the DF signal, i.e., blur the imaging data. Accordingly, different experimental setups and different theoretical DF retrieval algorithms have different internal rulers that dictate their phase and DF sensitivities. Therefore, we expect the retrieved DF images to differ to some extent, particularly for features at length scales near the threshold between being resolved and unresolved -- similar to the length of the method’s internal ruler.\\\\
The DF sensitivity of experimental techniques that employ a modulated X-ray beam, like single-grid and speckle-based imaging, partly depends on the characteristic size of the beam modulations relative to the size of the sample structures.  Modulation patterns with smaller feature sizes are more easily blurred than those with larger features, making them more sensitive to subtle DF effects. However, this also means they are more susceptible to saturation -- that is, the point at which the modulation is completely blurred out and the retrieved DF signal reaches its maximum. It is therefore essential to select an appropriate modulation size for the sample being imaged, striking a balance between high sensitivity and the need to preserve the reference modulations and avoid saturation. This can generally be determined during the initial stages of the imaging experiments by visually inspecting the reference pattern in the absence and presence of the sample. In our experiments, the modulator used for single-grid imaging had a $\qty{52}{\micro\meter}$ (8 pixels) period, and the sandpaper used for speckle-based imaging gave an effective speckle size of $\qty{18.3}{\micro\meter}$ (2.8 pixels). Consequently, the speckle-based imaging system was more sensitive to those sample structures that only weakly scatter the X-ray beam, compared to the single-grid system. Building on the ideas from the previous paragraph, although the single-grid imaging appeared to be more precise, the speckle-based system in our case was more sensitive due to the reference modulations being appropriately sized to probe smaller structures in the samples, such as the fibers comprising the four-rod sample. This helps explain why sharp structures are visible in the speckle-based-retrieved DF images but not in the single-grid-retrieved images -- for example, the fibers in the reed diffuser, toothpick, and the tree twig denoted by the yellow arrows in Figs.~\ref{fig:FourRod}c and d. In contrast to these modulation-based methods, the dual-energy propagation-based `self-reference' approach relies solely on the sample possessing resolvable texture and producing sufficiently different local X-ray scattering at two energies. If either of these conditions is not met, DF retrieval becomes challenging for the approach described by Ahlers \textit{et al.} One way to avoid this limitation is to deliberately introduce a pattern to the sample within the dual-energy approach \cite{ahlers2024x}.\\\\
DF retrieval also depends on the theoretical underpinnings of the algorithm employed. Global algorithms that operate at the entire image level and provide closed-form analytical solutions to the DF inverse problem account for the spatial flow of information across the image, are computationally efficient, but may rely on stronger simplifying assumptions. These assumptions are made to simplify the mathematics of the inverse problem. As a result, regions of the sample that violate them may yield inaccurate DF retrieval; for example, a sample comprising materials with significantly different refractive indices would violate the single-material assumption in the dual-energy propagation-based method. In contrast, local pixel-wise approaches are more general and make fewer assumptions about the sample. These approaches do not involve analytically solving an equation to address the inverse problem. Instead, they recover the DF by quantifying local, pixel-scale changes to a reference pattern. These local approaches typically consider contrast effects only within finite-sized analysis windows, without accounting for intensity flows across the entire image, as global approaches do. A trade-off is that spatial resolution is limited by the modulation period and/or the size of the analysis window, as a median filter of the same size must be applied during post-processing to suppress artefacts resembling the reference pattern introduced by the moving-window analysis. The single- and multi-speckle approaches investigated in this work are global DF retrieval methods. As evident from the images presented, the retrieved DF signals are smooth, with artefacts arising primarily from experimental factors -- for example, the speckled appearance of the retrieved DF in the single-speckle reconstructions, most clearly seen in Fig.~\ref{fig:FourRod}c. The dual-energy propagation-based approach involves a global calculation followed by a local one, and therefore exhibits characteristics of both global and local methods. Finally, the single-grid approach is entirely local; its resolution is limited by the size of the analysis window used for retrieval (which is typically lower-bounded by the grid period). It also has the longest computation time of the approaches investigated.
\begin{figure}[!t]
    \centering
    \includegraphics[width=\textwidth]{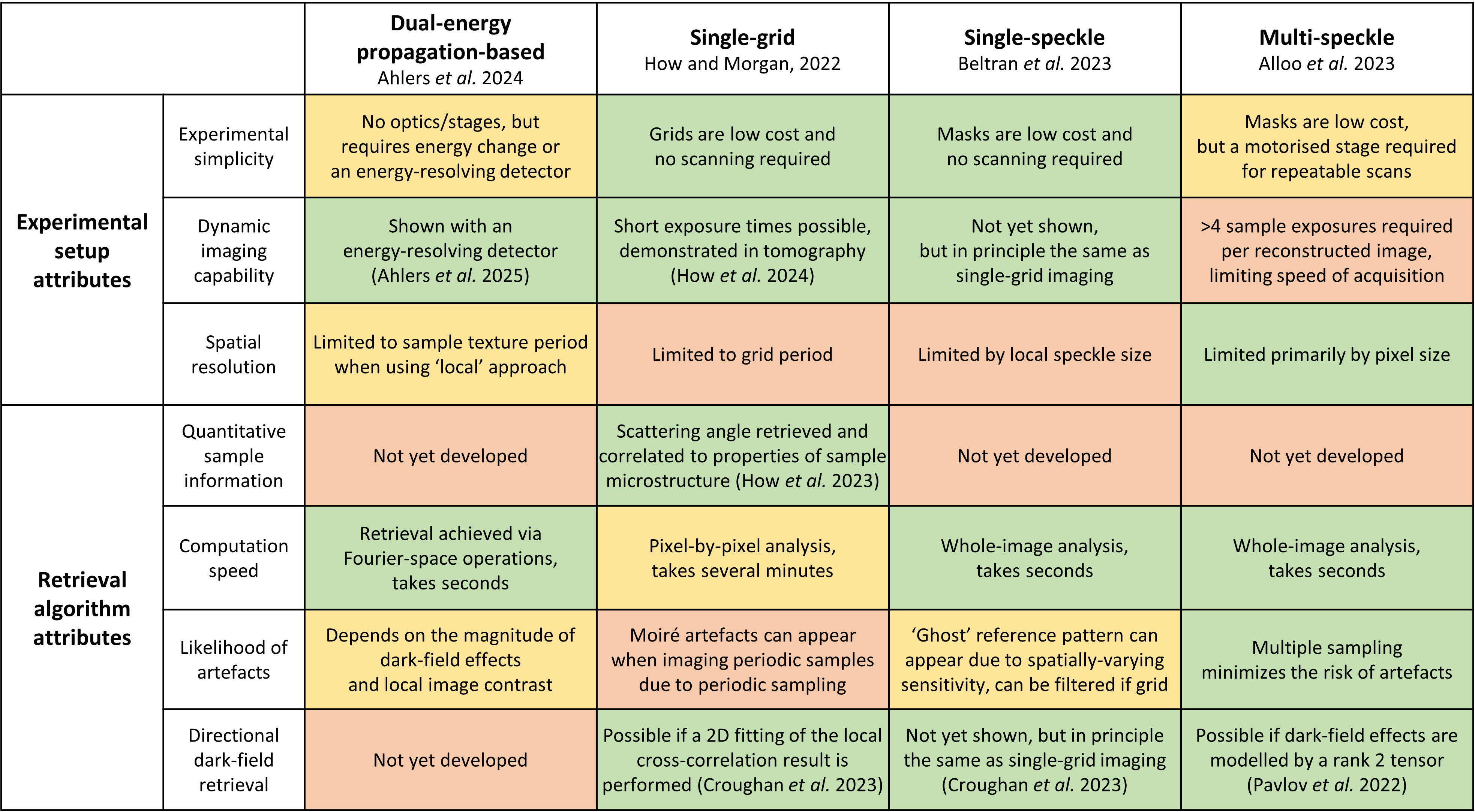} 
    \caption{Summary of the strengths and limitations of each DF-retrieval approach (top row), assessed against the desirable attributes listed down the left column.}
    \label{fig:Smiley}
\end{figure}
\subsection{Summary of approaches}\label{summary}
\noindent The strengths and limitations of the investigated DF approaches are summarized in the table shown in Fig.~\ref{fig:Smiley}. The DF approaches are listed along the top of the table, and desirable attributes relating to the experimental setup and DF retrieval algorithm are along the left-hand side. Each cell is filled with green, orange, or red, indicating how well a given approach satisfies the denoted attribute, where green indicates the best performance and red the weakest. This table serves as a practical guide to help DF imaging users select the most appropriate approach for their specific needs, from among the four methods compared in this paper. It is important to appreciate that there are ongoing developments in almost all methods, so this table may change in the coming years. Note also that the dual-energy approach uses the `local' method for both the results shown in this manuscript and the table here, but there also exists an alternative `global' method that is described in Ahlers \textit{et al.} \cite{ahlers2024x}. Finally, while each experimental data set is analyzed with just one retrieval algorithm, the single-grid and single-speckle retrieval algorithms can be applied to both sets of data, and the multi-speckle retrieval algorithm could also be applied to multiple grid images. For this reason, we separate experimental advantages from retrieval-algorithm advantages in the table.

\section{Conclusion}
In this work, we apply four emerging X-ray dark-field (DF) imaging techniques and retrieval approaches to a common set of samples to clarify how these different approaches capture and reconstruct the DF signal from the same material. Two samples containing several distinct materials each were imaged using propagation-based, single-grid, and speckle-based approaches. We found that all investigated DF approaches recover the primary (low-frequency) contribution to DF contrast similarly, however, discrepancies emerged at higher spatial frequencies (e.g., at edges) and in terms of potential reconstruction artefacts. These differences can be characterized in terms of each approach's DF \textit{range} and \textit{precision}, which collectively define \textit{internal rulers} that attribute the observed discrepancies and govern the structural features to which each method is sensitive. These rulers arise from both experimental design choices, e.g., grid period, sample-to-detector distance, pixel size, and the underlying assumptions of the retrieval algorithms. To support practical decision-making, we provide a summary table to help users select an appropriate DF technique based on their specific imaging goals. We hope this work contributes to a discourse around the complementarity of various DF imaging approaches---highlighting that each technique brings unique strengths and limitations, making them well-suited for different applications. Future studies inspired by this work could focus on quantifying the internal rulers of different DF methods and on better understanding the mechanisms of edge-induced DF contrast--how it manifests in experimental data and how it is interpreted by different retrieval algorithms, which remains an open question in DF imaging research.\\\\
\section*{Acknowledgements}
The authors thank beamline scientists Andrew Stevenson and Benedicta Arhatari for their support at the MicroCT beamline of the Australian Synchrotron, where all images in this paper were acquired under proposal 19663. We also thank Marie-Christine Zdora for her assistance with implementing a speckle-drift correction for the multi-speckle imaging data of the two samples. Kaye S. Morgan acknowledges support from the Australian Research Council (FT18010037 and DP230101327). Jannis N.~Ahlers and Michelle K.~Croughan acknowledge support from an Australian Government Research Training Program (RTP) Scholarship. Ying Ying How acknowledges support from a Monash Graduate Scholarship (MGS). Jannis N. Ahlers~acknowledges support from AINSE Ltd.~via a Postgraduate Research Award (PGRA). Samantha J.~Alloo and Michelle K.~Croughan acknowledge support from an AINSE Ltd.~Early Career Researcher Grant (ECRG).

\bibliographystyle{abbrv} 
\bibliography{references} 

\end{document}